\documentclass[a4paper]{article}

\usepackage{icrc2013}

\newcommand{\lsi}   {LS~I~+61~303}

\newcommand{\mnras} {Monthly Notices of the RAS}

\newcommand{\apj} {Astrophysical Journal}
\newcommand{\apjl} {Astrophysical Journal, Letters}
\newcommand{\aap} {Astronomy and Astrophysics}
\newcommand{\memsai} {Memorie della Societa Astronomica Italiana}

\title{Exploring high-energy processes in binary systems with the
Cherenkov Telescope Array}

\shorttitle{Binaries with the CTA}

\authors{
J.M. Paredes$^{1}$,
W. Bednarek$^{2}$,
P. Bordas$^{3}$,
V. Bosch-Ramon$^{4}$,
E. De Cea del Pozo$^{5}$,
G. Dubus$^{6}$,
S. Funk$^{7}$,
D. Hadasch$^{5}$,
D. Khangulyan$^{8}$,
S. Markoff$^{9}$,
J. Mold\'on$^{1,20}$,
P. Munar-Adrover$^{1}$,
S. Nagataki$^{10}$,
T. Naito$^{11}$,
M. de Naurois$^{12}$,
G. Pedaletti$^{5}$,
O. Reimer$^{13}$,
M. Rib\'o$^{1}$,
A. Szostek$^{14,15}$,
Y. Terada$^{16}$,
D.F. Torres$^{17,5}$,
V. Zabalza$^{1,18}$,
A.A. Zdziarski$^{19 }$,
for the CTA Consortium.
}

\afiliations{
$^1$ Departament d'Astronomia i Meteorologia, Institut de Ci\`encies del Cosmos (ICC), Universitat de Barcelona (IEEC-UB), Mart\'{\i} i Franqu\`es 1, E-08028, Barcelona, Spain \\
$^2$ Department of Astrophysics, University of L\'od\'z, ul. Pomorska 149/153, 90-236 L\'od\'z, Poland \\
$^3$ Institut f\"ur Astronomie und Astrophysik, Universit\"at T\"ubingen, Sand 1, 72076 T\"ubingen, Germany \\
$^4$ Dublin Institute for Advanced Studies, 31 Fitzwilliam Place, Dublin 2, Ireland \\
$^5$ Institut de Ci\`encies de l'Espai (IEEC-CSIC), 08193 Bellaterra, Spain \\
$^6$ UJF-Grenoble 1/CNRS-INSU, Institut de Plan\'etologie et d'Astrophysique de Grenoble (IPAG) UMR 5274, 38041 Grenoble, France \\
$^7$ W. W. Hansen Experimental Physics Laboratory, Kavli Institute for Particle Astrophysics and Cosmology, Department of Physics and SLAC National Accelerator Laboratory, Stanford University, Stanford, CA 94305, USA \\
$^8$ Institute of Space and Astronautical Science/JAXA, 3-1-1 Yoshinodai, Chuo-ku, Sagamihara, Kanagawa 252-5210, Japan \\
$^9$ Astronomical Institute �Anton Pannekoek�, University of Amsterdam, P.O. Box 94249, 1090 GE Amsterdam, The Netherlands \\
$^{10}$ Yukawa Institute for Theoretical Physics, Kyoto University, Kitashirakawa Oiwake-cho, Sakyo-ku, Kyoto 606-8502, Japan \\
$^{11}$ Faculty of Management Information,Yamanashi Gakuin University, Sakaori 2-4-5, Kofu-si, Yamanashi 400-8575, Japan \\
$^{12}$ Laboratoire Leprince-Ringuet, Ecole Polytechnique, CNRS/IN2P3, F-91128 Palaiseau,
France \\
$^{13}$ Institut f\"ur Astro- und Teilchenphysik, Leopold-Franzens-Universit\"at Innsbruck, A-6020 Innsbruck, Austria \\
$^{14}$ Kavli Institute for Particle Astrophysics and Cosmology, SLAC National Accelerator Laboratory, 2575 Sand Hill Road, Menlo Park, CA 94025, USA \\
$^{15}$ Astronomical Observatory, Jagiellonian University, Orla 171, 30-244 Krak\'ow, Poland \\
$^{16}$ Graduate School of Science and Engineering, Saitama University, 255 Simo-Ohkubo, Sakura-ku, Saitama City, Saitama 338-8570, Japan \\
$^{17}$ ICREA, 08010 Barcelona, Spain \\
$^{18}$ Max-Planck-Institut f\"ur Kernphysik, PO Box 103980, 69029 Heidelberg, Germany \\
$^{19}$ Centrum Astronomiczne im. M. Kopernika, Bartycka 18, 00-716 Warszawa, Poland \\
$^{20}$ ASTRON Netherlands Institute for Radio Astronomy
Oude Hoogeveensedijk 4, 7991 PD Dwingeloo, The Netherlands \\
}
\email{jmparedes@ub.edu}

\abstract{Several types of binary systems have been detected up to now at high and very high gamma-ray energies, including microquasars, young pulsars around massive stars and colliding wind binaries. The study of the sources already known, and of the new types of sources expected to be discovered with the unprecedented sensitivity of CTA, will allow us to qualitatively improve our knowledge on particle acceleration, emission and radiation reprocessing, and on the dynamics of flows and their magnetic fields. Here we present some examples of the capabilities of CTA to probe the flux and spectral changes that typically occur in these astrophysical sources, as well as to search for delays in correlated X-ray/TeV variability with CTA and satellites of the CTA era. Our results show that our knowledge of the high-energy physics in binary systems will significantly deepen with CTA.}


\begin{document}
\maketitle

%Begin a section.
\section{Introduction}
The detection of Very High Energy (VHE)  gamma rays ($E >$ 100 GeV) by the 
current Imaging Atmospheric Cherenkov Telescopes (IACT) from the binary 
systems PSR B1259$-$63 \cite{2005A&A...442....1A, 2009A&A...507..389A}, LS~5039 \cite{2005Sci...309..746A}, LS~I~+61~303 \cite{2006Sci...312.1771A, 2008ApJ...679.1427A} and HESS J0632$+$0757 \cite{2007A&A...469L...1A}, and the evidence of detection of a VHE flare in the black hole binary Cygnus X-1 \cite{2007ApJ...665L..51A}, provides a clear evidence of very efficient particle acceleration in binary systems containing compact objects (see e.g. \cite{2011arXiv1101.4843P}). In addition, the source 1FGL~J1018.6$-$5856 has been proposed to be a new gamma-ray binary \cite{2012Sci...335..189F} that could be associated with a H.E.S.S. source \cite{2010cosp...38.2803D}. Furthermore, there are other binary systems from which VHE emission is expected from a theoretical point of view \cite{2011MmSAI..82..182B}. 

Studying known and new compact binary systems at VHE is of great 
importance because their complexity allows us to probe physical processes 
that are still poorly understood. For instance, some of these systems are 
extremely efficient accelerating particles and their study could shed new light on, and eventually force a revision of, particle acceleration theory (see e.g. \cite{2008MNRAS.383..467K}). CTA will allow us to go beyond the current IACT's and offer a unique opportunity to perform a deeper analysis of the processes taking place in binary systems. Below, we report on examples of numerical simulations performed to show how the forthcoming CTA observatory \cite{2011ExA....32..193A} could fulfill the new scientific challenges. A more detailed study of the capability of CTA to probe the spectral, temporal and spatial behavior of gamma-ray binaries can be found in \cite{2013APh....43..301P}. Another work including some studies about these binaries is \cite{2013APh....43...81B}

\section{Testing the CTA capabilities} 
Several of the studies herein presented are meant as examples of the capability of CTA in comparison with current IACTs; thus we take here several of the already detected gamma-ray binaries, for which a well known set of assumptions can be established. We have used Monte Carlo simulations of the sensitivity of the array for several possible configurations in order to explore the capability of CTA to study  binary systems. In particular, we have conducted simulations using configurations\footnote{Arrays \texttt{E} and \texttt{I} are balanced layouts, in terms of the distribution of resources across the full CTA energy range. Array \texttt{B} is more focussed at low energies and \texttt{D} at high energies. \texttt{NB} is a higher energy focussed alternative to \texttt{NA}.} \texttt{B}, \texttt{D}, \texttt{E}, \texttt{I}, \texttt{NA} and \texttt{NB}, and subarrays \texttt{s4-2-120}, and \texttt{s9-2-120}, where the first number indicates the number of telescopes, the second indicates the type (1 for Small Size Telescope, 2 for Medium and 3 for Large), and the last number indicates the separation in meters. Details about the configuration and subarray characteristics can be found in \cite{2013APh....43..171B}.
%\cite{2013APh....43....1H}.

\subsection{CTA flux error reduction in known TeV sources}\label{error}
We studied the modulation of the photon index and the flux normalization with the orbital period for a source like  LS\,5039. The direct comparison of the errors of the H.E.S.S. and CTA measurements shows that observations with CTA can reduce the errors on the spectral parameters by a factor between 2 and 4.5, and that CTA could easily distinguish between spectra at different orbital phase bins.

\subsection{Short timescale flux variability}\label{fast}
We have studied the capabilities of CTA to detect short timescale flux and spectral variability from binaries that produce gamma rays. The perfect candidate for such studies is \lsi. A simultaneous multi-wavelength campaign on this source resulted in the discovery of correlated X-ray/VHE emission with the MAGIC IACT and the X-ray satellites \emph{XMM-Newton} and \emph{Swift} in the energy range 0.3--10 keV \cite{2009ApJ...706L..27A}. Under the assumption that the X-ray/VHE emission correlation found by \cite{2009ApJ...706L..27A} holds at shorter timescales, we took the X-ray light curve (LC) from a $\sim$100\,ks long \emph{Chandra} observation \cite{2010MNRAS.405.2206R} and computed the expected VHE emission from the source. 

Two feasible observation scenarios have been considered. In the first one, the full CTA \texttt{I} array is used to carry out a deep, short observation of \lsi.  In this case, the VHE counterpart of fast X-ray variability down to 1000\,s is clearly detected by CTA simulations, and also the X-ray/VHE correlation is recovered, with $r\ge0.9$, for flux variations of a factor 1.5.
In the second observation scenario, a ten nights campaign with three hours of observation per night is simulated using the continuous 100\,ks X-ray LC. This campaign would be done with the subarray \texttt{s9-2-120}, which is a subarray configuration similar to an expanded H.E.S.S. The longer observation time allows the campaign to probe a wider range of X-ray fluxes than the variation of a factor 1.5 used for the shorter, full CTA array observation discussed above.

To further explore the shortest time scales in which CTA can resolve a 
flare, we simulated a 20\,hrs event whose flux variation follows a Gaussian distribution and assuming the best-fit spectral shape reported by MAGIC for the Cygnus X-1 signal \cite{2007ApJ...665L..51A}. We find that the full CTA \texttt{I} array could clearly resolve a gaussian-shaped flare with a binning of 5 minutes for each data point. A 5 minute integration would result in a detection with a significance of 7 $\sigma$ at the assumed low state, and 25 $\sigma$ in high state, whereas with the sensitivity of MAGIC it is only possible to detect the peak of this flare. This clearly shows the advantadge in sensitivity that CTA will bring at very high energies with respect to the current IACTs.

As a conclusion from this simulation exercise, we see that the full CTA array will be a powerful tool to probe the fast flux variability of gamma-ray binaries, which could allow the characterization of the dynamical processes taking place in the emitting plasma. In addition, we have shown that subarray operations are very useful since they are able to provide us with novel results using only a fraction of the full array.
%They will prove invaluable not only in observational campaigns where the
%long-term variability is important, but also to monitor flaring sources such as
%gamma-ray binaries and AGNs.

\subsection{Sensitivity to spectral shape variations}
\begin{figure}[]
  \centering
  \includegraphics[width=0.48\textwidth]{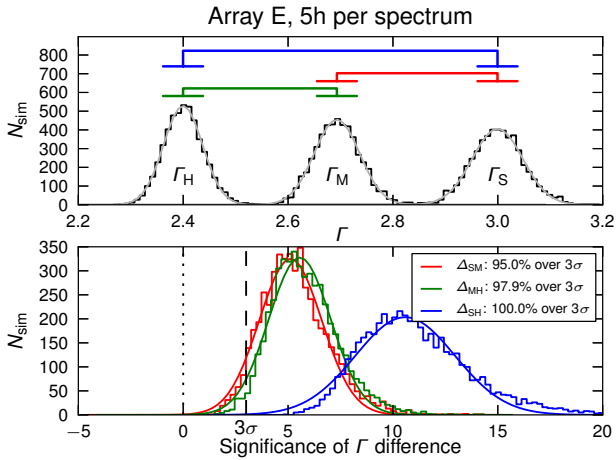}
  \caption{Results of 10\,000 realizations of the spectral observation of 
LS~I~+61~303 with the array \texttt{E} during 5\,h per spectral state. 
\emph{Top:} Distribution of recovered spectral indices for soft 
($\Gamma_\mathrm{S}=3.0$), mean ($\Gamma_\mathrm{M}=2.7$), and hard 
($\Gamma_\mathrm{H}=2.4$) spectra. \emph{Bottom:} Distribution of 
significances of the difference between each of the spectrum pairs. The 
legend indicates the fraction of realizations with a significance larger
than $3\sigma$.}
  \label{lsispecs}
 \end{figure}

To explore the power of CTA to distinguish between different spectral 
indices, we have considered a variation of the power-law VHE spectrum of \lsi\ within the statistical error obtained in the MAGIC campaign on September 2007 (see \cite{2009ApJ...706L..27A}) and tested the time required by CTA to significantly detect it.  We took the two extreme values of the measured spectral index (2.4 and 3.0) and obtained the simulated CTA spectra.  The result is that observing each of the spectral states for 3\,h (i.e., one night per state) with the full CTA \texttt{E} array provides a set of spectra from which the photon index variations can be clearly seen. We note that an increase in observation time to 5\,h per spectrum leads to a detection rate above 95\% for all the spectrum pairs (see Figure~\ref{lsispecs}). We also considered the possibility of longer observations with the subarray \texttt{s9-2-120}. In this case, for an observation of 10\,h for each of the spectral states, the spectral variation between the soft and the mean spectra is still significant at a level higher than $3\sigma$ for 82\% of the simulation realizations.

It is clear that CTA will be a powerful tool for the detection of spectral variations in gamma-ray binaries. The statistical errors we have obtained for these simulations are nearly an order of magnitude lower than the ones obtained with MAGIC in 2007, thus demonstrating the capability of CTA to deliver new and exciting science in the following years.

\subsection{Exploring the minimum detectable time delay between X-ray and TeV emission in gamma-ray binaries}

If there were a time delay between the non-thermal emission at different bands (e.g., X-ray and TeV) larger than the electron cooling time scale, the emission from these bands would most likely have an origin in different locations in the binary. The detection of such a delay has not been possible for the current generation of IACTs, so here we present a study of the capability of CTA to achieve this goal. Since CTA will operate together with a new generation of X-ray telescopes, such as the Japanese X-ray telescope {\it Astro-H}, we have done this study considering the capabilities of this new instrument.

Based on the short X-ray flares from \lsi\ detected in \cite{2010MNRAS.405.2206R}, we have simulated the LC detected by \emph{Astro-H}. Using the correlation between X-rays and VHE gamma rays found in \cite{2009ApJ...706L..27A} and the tools for simulating the CTA response, we generate the corresponding VHE LC of the flare as seen by CTA \texttt{I} array above 65 GeV. We have used a time binning of 600 seconds for both the CTA and the {\it Astro-H} LC, and studied positive delays of the TeV LC with respect to the X-ray LC in the range 0 to 2000 s in steps of 100 s. 
%The simulated X-ray fluxes and the TeV fluxes have a low correlation coefficient due to a loop structure induced by the delay.

\begin{figure}[]
  \centering
  \includegraphics[width=0.48\textwidth]{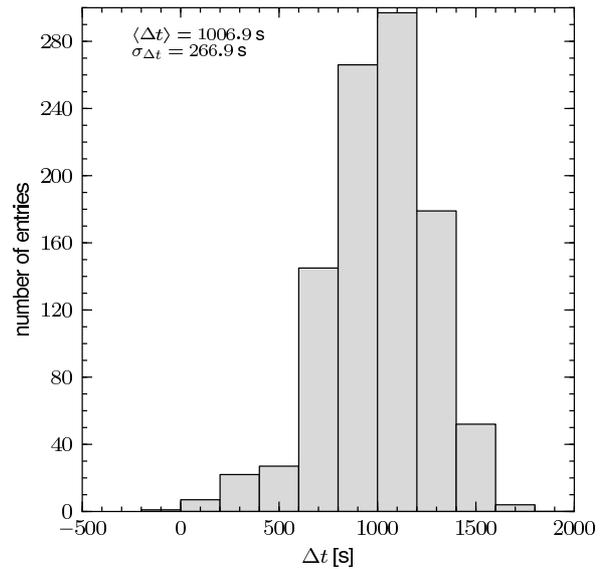}
  \caption{Distribution of the fitted peak of the ZDCF for the simulated light-curves for an introduced delay of 1000 s.
   A Gaussian fit gives the measured delay $\left<\Delta t\right>$ and its standard deviation $\sigma_{\Delta t}$.}
  \label{histo}
 \end{figure}

To clearly detect such delayed correlations, we have used the $z$-transformed Discrete Correlation Function (ZDCF), which determines 68\% confidence level intervals for the correlation coefficient for running values of the delay (see, e.g., \cite{1988ApJ...333..646E, 1997ASSL..218..163A}). Since the VHE LC is generated from the X-ray one, their errors are correlated and the scattering of one original (X-ray) LC affects the derived one (VHE). To correct this problem simulations have been performed: we started with 100 \emph{Astro-H} LC. With each of this 100 LC we produced 10 other X-ray LC by adding a Gaussian noise to the original ones, for each simulated delay. Then we simulated the corresponding VHE LC. In this way we obtained, for each delay, a sample of 1000 pairs of LC. For each pair of LC we have calculated the ZDCF. To evaluate how significant is the measurement of a delay using the ZDCF we have fitted Gaussian functions to the maxima of the ZDCF. At the end we have 1000 values of the peak for each simulated delay distributed around the real delay (see Figure~\ref{histo}). The measured delay is then calculated as the mean value of the distribution and its uncertainty is the standard deviation. Considering all possible uncertainties, delays of $\sim$1000 s can be significantly detected at a 3$\sigma$ confidence level in simultaneous LC obtained with  \emph{Astro-H} and CTA. These results are, to first order, independent of the duration of the short flares, as far as they last longer than the binning. Overall, these results indicate that CTA will allow us to localize and constrain the X-ray and TeV emitting regions of gamma-ray binaries and their properties.

\subsection{Exploring the collision of microquasar jets with the interstellar medium}

To simulate the CTA response to the observation of MQ jet/ISM interactions, we have used the theoretical predictions for a source with a jet power of $ 10^{38}$~erg~s$^{-1}$ and an age of $10^{5}$~yr embedded in a medium with particle density of $1$~cm$^{-3}$ \cite{2009A&A...497..325B}. Gamma-ray spectra with $\Gamma_{\rm ph} \sim 2.45$ and $\Gamma_{\rm ph} \sim 2.85$ are derived from current theoretical models for the leptonic and hadronic contribution, respectively \cite{2009A&A...497..325B, 2011MNRAS.410..978Z}. We studied the CTA performance in 50~h of observation time using the simulation tools for the \texttt{B} and \texttt{E} array configurations. The simulated flux is $\sim 1$\% of that of the Crab Nebula, although the steepening of the spectrum at high energies would make it difficult to detect the sources above a few TeV. 
 
Regarding the extension of the emission as seen in gamma rays, accelerated particles emitting at VHE do not have time to propagate to large distances since radiative cooling is very effective, and the emission will be mostly confined to the accelerator region. The emitter size may not largely exceed the width of the jet, $\sim 1$~pc, in the reverse shock region. In the case of the bow shock, although it may extend sideways for much larger distances, only the region around its apex ($\sim$~few~pc) will effectively accelerate particles up to the highest energies. The total angular size of the emission from a source located 3~kpc away may then be $\leq$~few arcmin, and CTA would image a point-like source with only a marginal extension roughly perpendicular to the jet direction. 

\subsection{Exploring the colliding winds of massive star binary systems}

\begin{figure}[]
  \centering
  \includegraphics[width=0.5\textwidth]{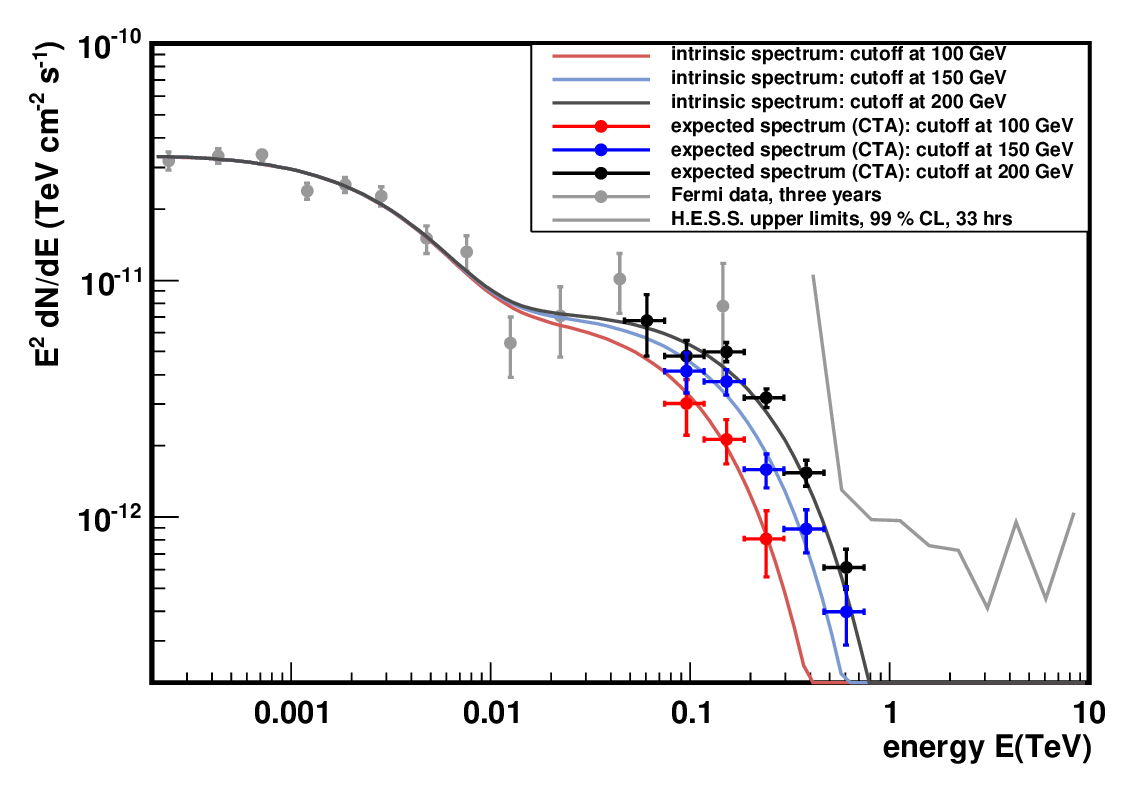}
  \caption{Intrinsic and CTA simulated spectra of Eta Carinae after 10 hours of observation time with high energy cutoffs at 100, 150 and 200\,GeV.}
  \label{etacarsimul}
 \end{figure}

We studied colliding wind binaries like Eta Carinae by means of simulations of the response of CTA. We based our simulations on the measurements of the energy spectrum of Eta Carinae by the {\em Fermi}/LAT \cite{2012A&A...544A..98R} and the upper limits derived by the H.E.S.S. Collaboration \cite{2012MNRAS.424..128H}. The spectrum between 0.1 and 100\,GeV is best fit by a power law with an exponential cutoff plus an additional power law at high energies. 
%In Figure \ref{etacarsimul} we show the Fermi/LAT data points and the H.E.S.S. upper limits in gray. From these measurements, it seems that there must be a cutoff in the spectrum at high energies. 
For our simulations we assume exponential cutoffs  at $E={100, 150, 200}$ GeV and test how well CTA could detect those. We produced simulations at increasing observation times in order to study the minimal time required to detect the source and to get a meaningful spectra with such CTA observations. In Figure~\ref{etacarsimul}, we show simulated energy spectra, for 10 hours of observation and different cutoffs, as they would be measured by CTA. The minimum observation time needed to significantly determine the cutoff energy, i.e. to distinguish between a simple power law and a cutoff power law, is established using the likelihood ratio test for the two hypothesis. From our simulations, we can conclude that CTA observation times of $>$15\,h are necessary to make a meaningful physical interpretation and modeling, whereas $>$20\,h allow the precise measure of the spectral energy cutoff. 

\section{Conclusions}

The sensitivity of CTA   will lead to a very good sampling of LC and spectra on very short timescales. It will allow as well long source monitoring using subarrays, still with a sensitivity 2--3 times better than any previous instrument operating at VHE energies. CTA will also reduce by a factor of a few the errors in the determination of fluxes and spectral indexes. Furthermore, the high sensitivity and good angular resolution will allow for imaging of possible extended emission in gamma-ray binaries, expected at the termination of the generated outflows. The low energy threshold will also permit to study the maximum particle energy achievable in massive star binaries, trace the effects of electromagnetic cascades in the spectra of gamma-ray binaries, or catch the most luminous part of the spectrum in some sources. Finally, under CTA the population of gamma-ray binaries (and their different subclasses) may easily grow by one order of magnitude, which will imply a strong improvement when looking for patterns and trends, to trace the physical mechanisms behind the non-thermal activity in these sources. For all this, CTA
%, either in highly sensitive observations of the whole array, or under the more suitable for monitoring subarray mode, 
will be a tool to obtain the required phenomenological information for deep and accurate modeling of gamma-ray binaries.  This can mean a qualitative jump in our physical knowledge of high-energy phenomena in the Galaxy.

\vspace*{0.2cm}
\footnotesize{{\bf Acknowledgment: }{We gratefully acknowledge support from the agencies and organizations listed in this page: http://www.cta-observatory.org/?q=node/22. J.M.P., V.B-R, P.M.-A., J.M., M.R. and V.Z. acknowledge support by DGI of the Spanish Ministerio de Econom\'{\i}a y Competitividad (MINECO) under grants AYA2010-21782-C03-01 and FPA2010-22056-C06-02. J.M.P. acknowledges financial support from ICREA Academia. J.M. acknowledges support by MINECO under grant BES-2008-004564. M.R. acknowledges financial support from MINECO and European Social Funds through a Ram\'on y Cajal fellowship.  V.Z. was supported by the Spanish MEC through FPU grant AP2006-00077. P.B. has been supported by grant DLR 50 OG 0601 during
this work. V.B.-R. acknowledges the support of the European Community under a Marie Curie Intra-European fellowship and financial support from MINECO through a Ram\'on y Cajal fellowship. D.H., G.P and D.F.T. acknowledge support from the Ministry of Science and the Generalitat de Catalunya, through the grants AYA2009-07391 and SGR2009-811, as well as by ASPERA-EU through grant EUI-2009-04072. GD acknowledges support from the European Community via contract ERC-StG-200911. The research leading to these results has received funding from the
European Union's Seventh Framework Programme ([FP7/2007-2013]
[FP7/2007-2011]) under grant agreement n¡ 262053.
}}

\end{document}